\pdfoutput=1
\documentclass[debug]{rmaa}

\newcommand{\Msun}{\mbox{M$_{\odot}$}}
\newcommand{\Rsun}{\mbox{R$_{\odot}$}}

\newcommand{\kms}{\mbox{km\,s$^{-1}$}}

\usepackage{supertabular}

\usepackage{paralist}

\usepackage{psfrag,color}

\usepackage[latin1]{inputenc}

\title{KIC\,9451096: Magnetic activity, flares and differential rotation} 

\author{
  O. \"Ozdarcan,\altaffilmark{1} 
  E. Yolda\c{s},\altaffilmark{1}
  and H. A. Dal\altaffilmark{1}}

\altaffiltext{1}{Ege University, Science Faculty, Department of Astronomy and Space Sciences, 
35100 Bornova, \.{I}zmir, Turkey.}

\shortauthor{\"Ozdarcan, Yolda\c{s}, \& Dal}
\shorttitle{KIC\,9451096}

\fulladdresses{
\item O. \"Ozdarcan, E. Yolda\c{s} and H. A. Dal: Ege University, Science Faculty, 
Department of Astronomy and Space Sciences, 35100 Bornova, \.{I}zmir, Turkey.
}

\listofauthors{O. \"Ozdarcan, E. Yolda\c{s}, \& H. A. Dal}
\indexauthor{\"Ozdarcan, O.}
\indexauthor{Yolda\c{s}, E.}
\indexauthor{Dal, H. A.}

\abstract{We present spectroscopic and photometric analysis of KIC\,9451096, where the latter is
based on very high precision long cadence photometry obtained by $Kepler$ space craft. Combined
spectroscopic and photometric modeling show that the system is a detached eclipsing binary in a 
circular orbit and composed of F5V + K2V components. Subtracting the best--fit light curve model 
from whole long cadence data reveals additional low (mmag) amplitude light variation in time and 
occasional flares, suggesting low, but still remarkable level of magnetic spot activity on the
K2V component. Analyzing rotational modulation of light curve residuals enables us to estimate
differential rotation coefficient of the K2V component as $k = 0.069 \pm 0.008$, which is 3 times weaker 
compared with the solar value of $k = 0.19$, assuming a solar type differential rotation. 
We find stellar flare activity frequency for K2V component as 0.000368411 $h^{-1}$ indicating low 
magnetic activity level.}

\resumen{}

\addkeyword{stars: activity}
\addkeyword{(stars:) binaries: eclipsing}
\addkeyword{stars: flare}
\addkeyword{stars: fundamental parameters}
\addkeyword{stars: individual (KIC\,9451096)}

\begin{document}
% Typeset article header
\maketitle

\section{Introduction}\label{S1}

Although the primary aim of $Kepler$ mission is to detect transiting planets by obtaining very high 
precision photometric measurements, it provides further benefits, especially in terms of clear and 
reliable determination of very small amplitude light variation on eclipsing and intrinsic variable 
stars. About 150\,000 targets have been observed in the mission, and apart from the exoplanets, which 
is the main purpose of the mission, numerous variable stars have been discovered. Unprecedented 
precision of $Kepler$ photometry clearly revealed low amplitude (mmag) light variations, which is 
used in analysis of stellar flares, spot activity and differential rotation 
\citep{Balona_2015MNRAS, Balona_2016MNRAS, Reinhold_Reiners_difrot_2013A&A, Reinhold_2013b_A&A}. 
Among these variable stars, 2876 eclipsing binary stars have been discovered \citep{Prsa_et_al_2011, 
Slawson_et_al_2011}. Careful light curve modeling of the binaries with cool components 
($T_{eff} < 6500$ K) revealed rotational modulation of light curves and flares in model 
residuals. KIC\,09641031 \citep{Yol16}, KIC\,09761199 \citep{Yol17} and KIC\,2557430 \citep{Kam17}, 
GJ\,1243, GJ\,1245A and B \citep{Hawley_et_al_2014ApJ}, KIC\,2300039, KIC\,4671547 \citep{Balona_2015MNRAS} 
are such stars.

The analyses of the patterns of magnetic activity exhibiting by these stars reveals some clues about 
their evolutionary stages. Although there are several indicators found in these analyses for the evolutionary stage, 
two of them are the energy spectra defined by \citet{Gershberg_1972Ap&SS} and flare frequencies described by 
\citet{Ishida_1991Ap&SS}. Both of them have been computed especially from the 1970's to the 1980's in order to figure out 
the magnetic activity levels for the stars, which the flares are detected from. In 1990's, \citep{Leto_1997A&A} examined 
the flare frequency variation of EV Lac, a well-known UV Ceti type star. There are a few studies, in which the activity levels of 
three magnetic active stars discovered in Kepler Mission are discussed depending on their flare frequencies, recently published in the literature. 
\citet{Yol16} detected 240 flares from KIC\,09641031, and \citet{Yol17} detected 94 flares from KIC\,09761199. In addition, 
\citet{Kam17} detected 69 flares from KIC\,2557430. \citet{Yol16} derived the One Phase Exponential Association 
(hereafter OPEA) model, and the flare frequency $N_{1}$ was found to be 0.41632 $h^{-1}$ for 
KIC\,09641031. \citet{Yol17} computed $N_{1}$ as 0.01351 $h^{-1}$ over 69 flares for KIC\,09761199. 
However, an interesting situation occurs in case of KIC\,2557430. \citet{Kam17} find that some of the 
flares detected from KIC\,2557430 come from a third body, which is unclear whether it is a component 
in the system or an undetected light source from background. Depending on the 
OPEA model derived from 69 flares, \citet{Kam17} reveal that 40 flares (called Group 1) of them 
come from the secondary component, while 29 flares (called Group 2) come from a third body. 
They computed the flare frequency $N_{1}$ as 0.02726 $h^{-1}$ for Group 1 and 0.01977 $h^{-1}$ for 
Group 2. As it is discussed by \citet{Yol16} and \citet{Ger05}, the flare frequency is one of the 
parameter which is an indicator about the nature of the flare mechanism that is in porgress on the 
stellar atmosphere. Apart from the classical parameters described by \citet{Ger05}, 
\citet{Dal_Evren_2010AJ, Dal_Evren_2011AJ} have also described some new parameters derived from the 
OPEA models in order to determine the flare process running on the stellar surface.

Continuous photometry of variable single stars discovered in the scope of $Kepler$ 
enabled to trace photometric period variation as a proxy of differential rotation via
Fourier transform \citep[see, e.g.][]{Reinhold_et_al_difrot_2013A&A, Reinhold_Reiners_difrot_2013A&A}. 
However, Fourier transform may not perfectly work in case of eclipsing binaries, where the 
amplitude of rotational modulation of star spots is usually embedded into the relatively large 
amplitude light variations caused by eclipses and break of spherical symmetry of the binary 
components. Furthermore, insufficient representation of light curve models, especially around
mid-eclipse phases, may require discarding of data around those phases and causes regular gaps in
light curve, which would lead to unwanted alias period and harmonics. In this case, alternative
methods could be adopted to trace photometric period variation, such as $O-C$ diagram based on 
minimum times of rotationally modulated light curves \citep[see, e.g.][]{V2075Cyg_Orkun_2010AN}.

In case of eclipsing binary stars, additional intrinsic variations may not be determined at first 
look, due to the reasons explained above. KIC\,9451096 is such an eclipsing binary in 
$Kepler$ eclipsing binary catalog\footnote{http://keplerebs.villanova.edu/}
\citep{Prsa_et_al_2011, Slawson_et_al_2011} with a short period, and with a confirmed third body 
\citep{Borkovits_2016MNRAS_To_P}. Beyond the properties provided by the catalog, such as morphology 
and eclipse depths, \citet{Armstrong_et_al_2014} provided physical information, estimated from 
spectral energy distribution based on photometric measurements. They estimated the effective temperature 
of the components of KIC\,9451096 as 7166 K and 5729 K for the primary and the secondary component, 
respectively.

In this study, we carry out photometric and spectroscopic analysis of KIC\,9451096, based on $Kepler$ 
photometry and optical spectroscopic observations with intermediate resolution ,described in 
Section~\ref{S2}. Section~\ref{S3} comprises spectroscopic and photometric modeling of the system, 
and the analysis of out--of--eclipse variations. In the final section, we summarize and discuss our 
findings.

\section{Observations and data reductions}\label{S2}

\subsection{$Kepler$ photometry}\label{S2.1}

Photometric data obtained by $Kepler$ spacecraft cover a broad wavelength 
range between 4100\,\AA~and 9100\,\AA, which has advantage of collecting much 
more photons in a single exposure and reaching sub-milimag precision, but also 
has disadvantage of having no "true" photometric filter, hence no photometric 
color information. There are two types of photometric data having different exposure 
times. These are short cadence data (having exposure time of 58.89 seconds) and 
long cadence data (having exposure time of 29.4 minutes). 
In this study we use long cadence data of KIC\,9451096 obtained from $Kepler$ 
eclipsing binary catalog. The catalog provides detrended and normalized intensities,
which is obtained by application of procedures described by 
\citet{Slawson_et_al_2011} and \citet{Prsa_et_al_2011}. The whole data covers $\sim$4 years 
of time span with 65\,307 data points in total. MAST archive reports 0.9\% 
contamination level in the measurements, practically indicating no additional light 
contribution to the measured fluxes of KIC\,9451096.

\subsection{Spectroscopy}\label{S2.2}

We obtained optical spectra of KIC\,9451096 by 1.5 m Russian -- Turkish 
telescope equipped with Turkish Faint Object Spectrograph Camera (TFOSC) at Tubitak National
Observatory\footnote{http://www.tug.tubitak.gov.tr/rtt150\textunderscore tfosc.php}.
TFOSC enables one to obtain intermediate resolution optical spectra in \'echelle mode. 
In our case, the instrumental setup provides actual resolution of R = $\lambda/\Delta\lambda$ 
$\sim$ 2500 around 6500\,\AA, and observed spectra covers usable wavelength range between 
3900--9100\,\AA~ in 11 \'echelle orders. Back illuminated 2048 $\times$ 2048 pixels CCD camera,
which has pixel size of 15 $\times$ 15 $\mu m^{2}$, was used to record spectra.

We obtained ten optical spectra of KIC\,9451096 between 2014 and 2016 observing 
seasons. In order to obtain enough signal, we used 3600 s of exposure time for each observation. 
Estimated signal--to--noise ratio (SNR) of observed spectra is mostly between 80--100, 
except a few case, where the SNR is around 50. SNR estimation is based on photon statistic. 
Together with the target star, we also obtained high SNR optical spectrum of HD\,225239 
(G2V, $v_{r} = 4.80$ \kms) and $\iota$\,Psc (HD\,222368, F7V, $v_{r} = 5.656$ \kms), and adopted
them as radial velocity and spectroscopic comparison templates.

We reduce all observations by using standard IRAF\footnote{The Image Reduction and Analysis 
Facility is hosted by the National Optical Astronomy Observatories in Tucson, Arizona at URL 
iraf.noao.edu.} packages and tasks. Typical reduction procedure starts with obtaining master bias
frame from nightly taken several bias frames and subtracting master bias frame from all object, 
calibration lamp (Fe-Ar spectra in our case) and halogen lamp frames. Then bias corrected halogen 
frames are combined together to form average halogen frame and this average frame is normalized to
the unity to produce normalized master flat frame. After that, all target and calibration lamp 
spectra are divided by the normalized flat field frame. Next, cosmic rays removal and scattered 
light corrections are applied to bias and flat corrected frames. At the end of these steps, reduced
frames are obtained and these frames are used for extraction of spectra. In the final steps, Fe-Ar 
frames are used for wavelength calibration of extracted spectra and wavelength calibrated spectra 
are normalized to the unity by using cubic spline functions.

\section{Analysis}\label{S3}

\subsection{Radial velocities and spectroscopic orbit}\label{S3.1}

\begin{figure}[!htb]
\centering
{\includegraphics[angle=0,scale=0.80,clip=true]{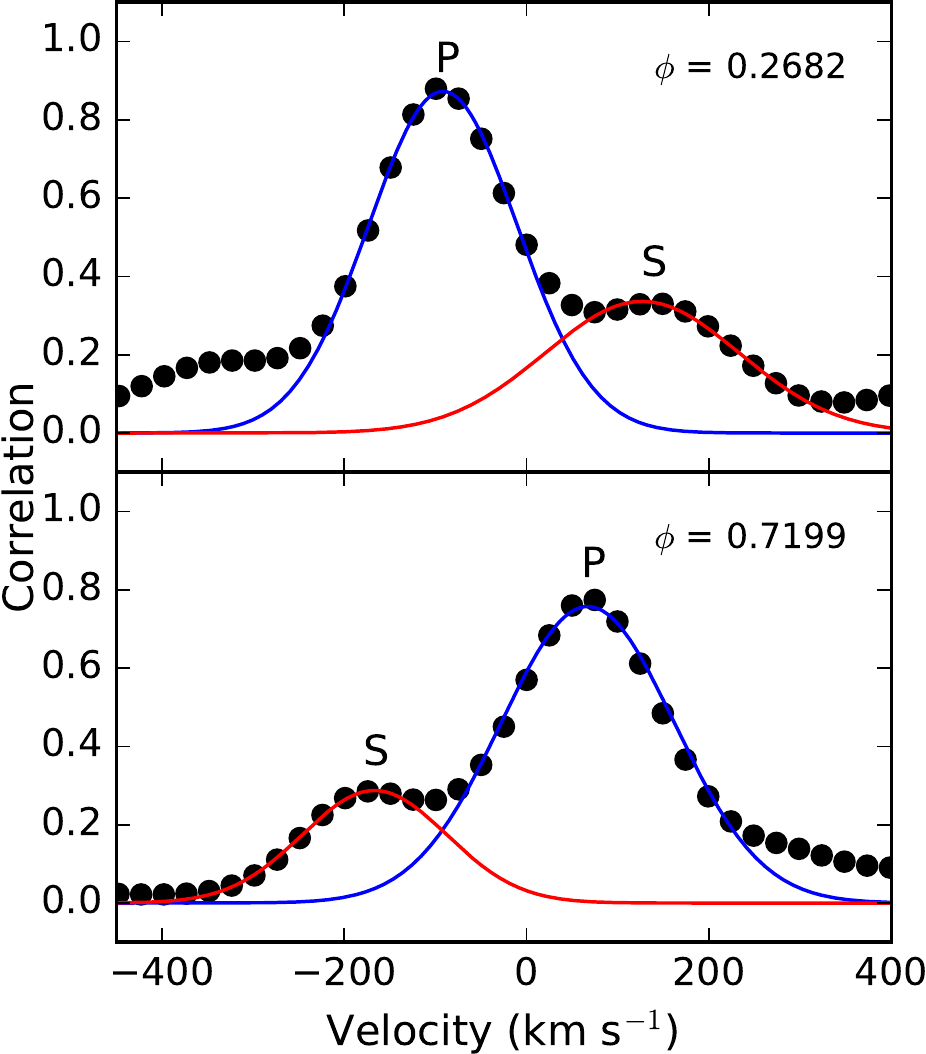}}
\caption{Cross-correlation functions of two spectra obtained in orbital quadratures. 
The letter $\phi$ denotes corresponding orbital phase. P and S indicate the primary 
component and the secondary component, respectively.}
\label{F1}
\end{figure}

The first step of our analysis is to determine radial velocities of the components and
spectroscopic orbit of the system. We cross-correlate each observed spectrum of KIC\,9451096
with spectra of template stars HD\,225239 and $\iota$\,Psc, as described in
\citet{fxcor_Tonry_Davis_1979}. In practice we use $fxcor$ task in IRAF environment. 
We achieve better cross-correlation signals (especially for the weak secondary component) when we 
use HD\,225239 as template, thus we determine all radial velocities with respect to the HD\,225239
spectrum. We obtain acceptable cross-correlation signals of both components in \'echelle orders 
5 and 6, which cover wavelength range between 4900--5700\,\AA\@. Figure~\ref{F1} shows 
cross-correlation functions of two spectra obtained around orbital quadratures.

%################ Figure 1 - CCFs

We list observational log and measured radial velocities of the components in Table~\ref{T1}. 
Note that we use ephemeris and period given by \citep{Borkovits_2016MNRAS_To_P} and listed in 
Table~\ref{T2} to calculate orbital phases and for further analysis.

% ----------------------- T1
\begin{table}
\setlength{\tabcolsep}{3pt}
\small
\caption{Log of spectroscopic observations together with measured radial velocities and their
corresponding standard errors ($\sigma$) in \kms.}\label{T1}
\begin{center}
\begin{tabular}{cccrrrr}
\hline\noalign{\smallskip}
      HJD    & Orbital  & Exposure & \multicolumn{2}{c}{Primary} &  \multicolumn{2}{c}{Secondary} \\
(24 00000+)  &  Phase   & time (s) & V$_{r}$ & $\sigma$ & V$_{r}$ & $\sigma$  \\
\hline\noalign{\smallskip}
56842.5435   &   0.7794   &   3600   &   91.4   &   8.2   &  -152.5   &   36.9   \\   %S/N @ 5500 83 A   
56844.4052   &   0.2682   &   3600   &  -79.9   &   6.3   &   151.9   &   39.1   \\   %S/N @ 5500 105 A   
56844.4479   &   0.3024   &   3600   &  -74.4   &   6.6   &   155.0   &   37.2   \\   %S/N @ 5500 88 A   
56889.4315   &   0.2781   &   3600   &  -77.1   &   5.7   &   148.1   &   40.0   \\   %S/N @ 5500 73 A   
56890.2958   &   0.9693   &   3600   &   14.5   &   5.0   &    ---    &    ---   \\   %S/N @ 5500 98 A   
57591.4532   &   0.7199   &   3600   &   88.5   &   7.2   &  -153.3   &   32.0   \\   %S/N @ 5500 48 A   
57601.4386   &   0.7058   &   3600   &   88.7   &   5.4   &  -149.8   &   32.1   \\   %S/N @ 5500 55 A   
57616.4778   &   0.7333   &   3600   &   86.0   &   4.3   &  -145.2   &   38.7   \\   %S/N @ 5500 55 A   
57617.5188   &   0.5659   &   3600   &   31.0   &   5.8   &    ---    &    ---   \\   %S/N @ 5500 79 A   
57672.3009   &   0.3779   &   3600   &  -54.8   &   5.1   &   111.1   &   47.9   \\   %S/N @ 5500 86 A   
\noalign{\smallskip}\hline
\end{tabular}
\end{center}
\end{table}

\begin{figure}[!htb]
\centering
{\includegraphics[angle=0,scale=0.55,clip=true]{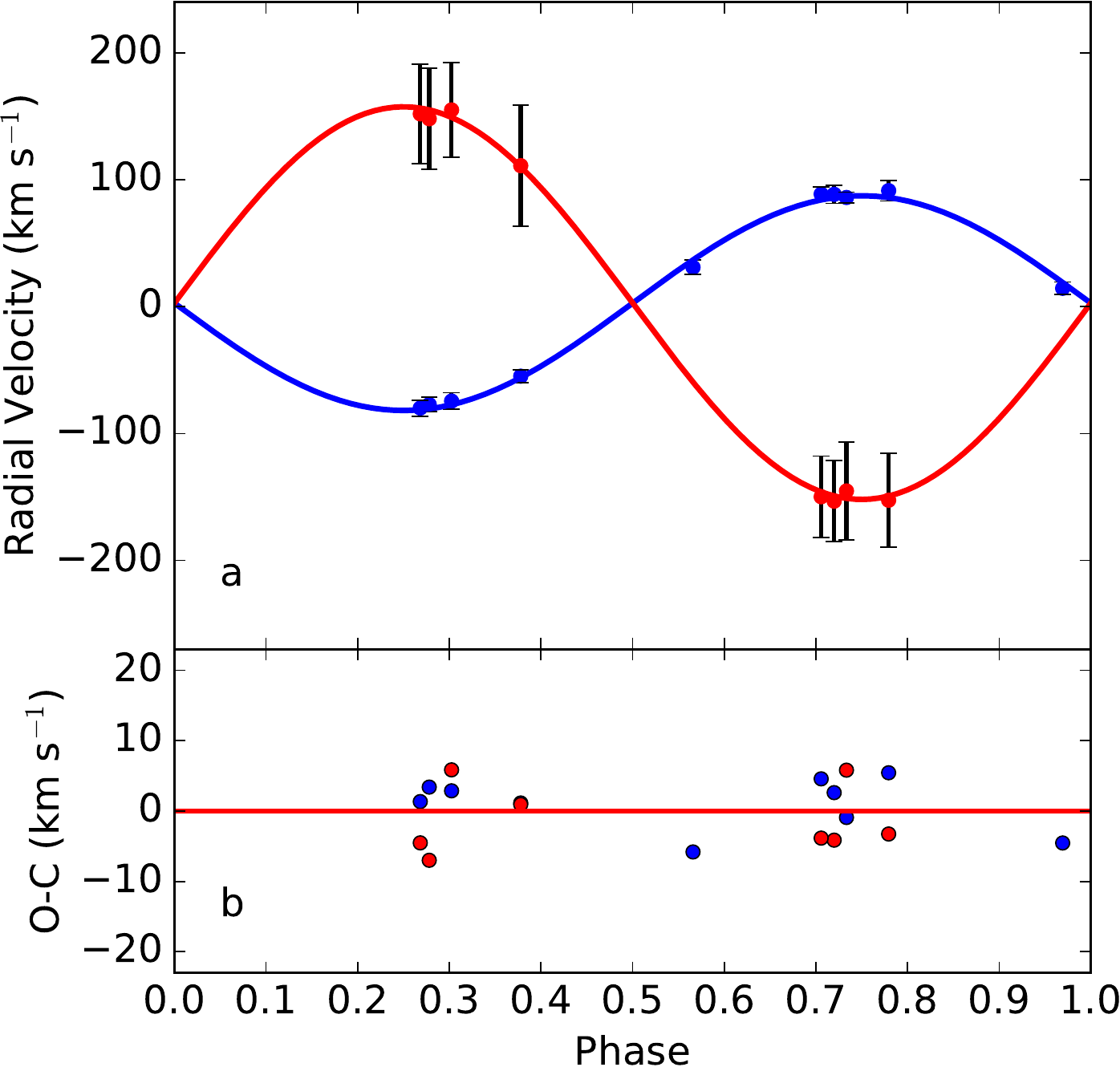}}
\caption{\textbf{a)} Observed radial velocities of the primary and the secondary (blue and red filled
circles, respectively), and their corresponding theoretical representations (blue and red curve).
\textbf{b)} Residuals from theoretical solution.}
\label{F2}
\end{figure}

We achieve reasonable solution for spectroscopic orbit under non-eccentric orbit assumption,
where the eccentricity is zero and the longitude of periastron is undefined. We check this assumption
by inspecting $Kepler$ light curve of the system, where we observe deeper and shallower eclipses at
0.0 and 0.5 orbital phases, respectively, indicating circular orbit (see Section~\ref{S3.3}, Figure~
\ref{F4}). In order to reach the final spectroscopic orbital solution, we prepare a simple script
written in python language, which applies Markov chain Monte Carlo simulations to the measured radial
velocities, considering their measured errors. We list the final spectroscopic orbital elements in 
Table~\ref{T2} and plot measured radial velocities, their observational errors, theoretical spectroscopic 
orbit and residuals from the solution in Figure~\ref{F2}.

% ----------------------- T2
\begin{table}
\caption{Spectroscopic orbital elements of KIC\,9451096. $M{_1}$ and $M{_2}$
denote the masses of the primary and the secondary component, respectively, while $M$ shows the total
mass of the system.}\label{T2}
\begin{center}
\begin{tabular}{cc}
\hline\noalign{\smallskip}
Parameter & Value \\
\hline\noalign{\smallskip}
$P_{\rm orb}$ (day)     &    1.25039069 (fixed)  \\
$T_{\rm 0}$ (HJD24 00000+)   &  54954.72942 (fixed)  \\
$\gamma$ (\kms)          &    2.8$\pm$0.5     \\
$K_{1}$ (\kms)           &     84.1$\pm$2.3     \\
$K_{2}$ (\kms)           &    153.2$\pm$14.6     \\
$e$                      &      0 (fixed)       \\
$a\sin i$ (\Rsun)        &     5.92$\pm$0.35    \\
$M\sin^{3} i$ (\Msun)    &    1.79$\pm$0.25  \\
Mass ratio ($q = M{_2}/M{_1}$)         &     0.55$\pm$0.05    \\
rms1 (\kms )           &         3.7        \\
rms2 (\kms )           &         4.9         \\
\noalign{\smallskip}\hline
\end{tabular}
\end{center}
\end{table}

%################ Figure 2 - RVs

\subsection{Spectral type}\label{S3.2}

We rely on our intermediate resolution TFOSC optical spectra to determine the spectral type of the
components. Most of our spectra correspond to the phases around orbital quadratures, where we observe
the signal of two components separated. However, there are two spectra obtained at phases close to 
the eclipses, where two components can not be resolved separately. One of these spectra corresponds to 
$\sim$0.56 orbital phase (see Table~\ref{T1}), where we can not observe the radial velocity signal 
of the secondary component in cross-correlation. Even in the orbital quadratures, cross-correlation
signal of the secondary component is considerably weak compared to the primary component, indicating 
a very small light contribution from the secondary component to the total light of the system. Our 
preliminary light curve analysis shows that the contribution of the secondary component to the total
light does not exceed $\sim$10\%. In this case, the signal from the secondary component becomes
almost negligible in the resolution of our observed spectrum at $\sim$0.56 orbital phase, therefore
we assume that we only observe the spectrum of the primary component and adopt this spectrum as
reference spectrum for the primary component. We confirm this assumption by calculating composite 
spectrum of the binary via final parameters of the components (see Section~\ref{S3.3}), where we 
observe that the contribution of the secondary component affects the theoretical composite spectrum 
less than 2\%for wavelength range of 4900-5700\,\AA\@. We refrain from performing detailed analysis 
with spectral disentangling. Future studies could take advantage of this technique and derive 
atmospheric parameters of the secondary.

We first compare the reference spectrum with the template spectra of HD\,225239 and $\iota$\,Psc.
We observe that $\iota$\,Psc spectrum provides closer match to the reference spectrum but also
indicates earlier spectral type and slightly lower metal abundances for the primary component. 
At that point, we switch to the spectrum synthesizing method. We use the latest version of python
framework $iSpec$ \citep{iSpec_Cuaresma_2014A&A} which enables practical and quick calculation of 
a synthetic spectrum with a given set of atmospheric parameters via different radiative transfer codes. 
Among these codes we adopt SPECTRUM\footnote{http://www.appstate.edu/$\sim$grayro/spectrum/
spectrum.html} code \citep{spectrum_gray_1994}, together with ATLAS-9 \citep{ATLAS9_castelli_2004}
model atmospheres and actual line list from the third version of the Vienna atomic line database
($VALD3$) \citep{VALD3_Ryabchikova_2015}.

Considering spectral type of $\iota$\,Psc, we synthesize spectra for the effective temperatures 
between 6000 K and 7000 K in steps of 250 K, and metallicity values ([Fe/H]) between 
$-$1.0 and 0.0 in steps of 0.5. For all synthetic spectra we fix the gravity (log $g$)
to 4.15, which we precisely calculate in light curve modeling (see Section~\ref{S3.3}).
Since we do not have high resolution spectrum, we fix the microturbulence velocity to 2 \kms.
We convolve all calculated spectra with a proper Gaussian line spread function in order to 
degrade their resolution to the resolution of TFOSC spectra. Instrumental broadening in TFOSC
spectra is 2.2 \,\AA\@, corresponding 119 \kms\@ for wavelengths around 5500\,\AA\@. Estimated 
projected rotational velocities of the components are 62 \kms\@ and 36 \kms\@ for the primary and 
the secondary component respectively (see Section~\ref{S3.3}). Since instrumental broadening is 
the most dominant broadening source in observed spectra, we do not consider rotational broadening 
and other line broadening mechanisms.
 
\begin{figure}[!htb]
\centering
{\includegraphics[angle=0,scale=0.80,clip=true]{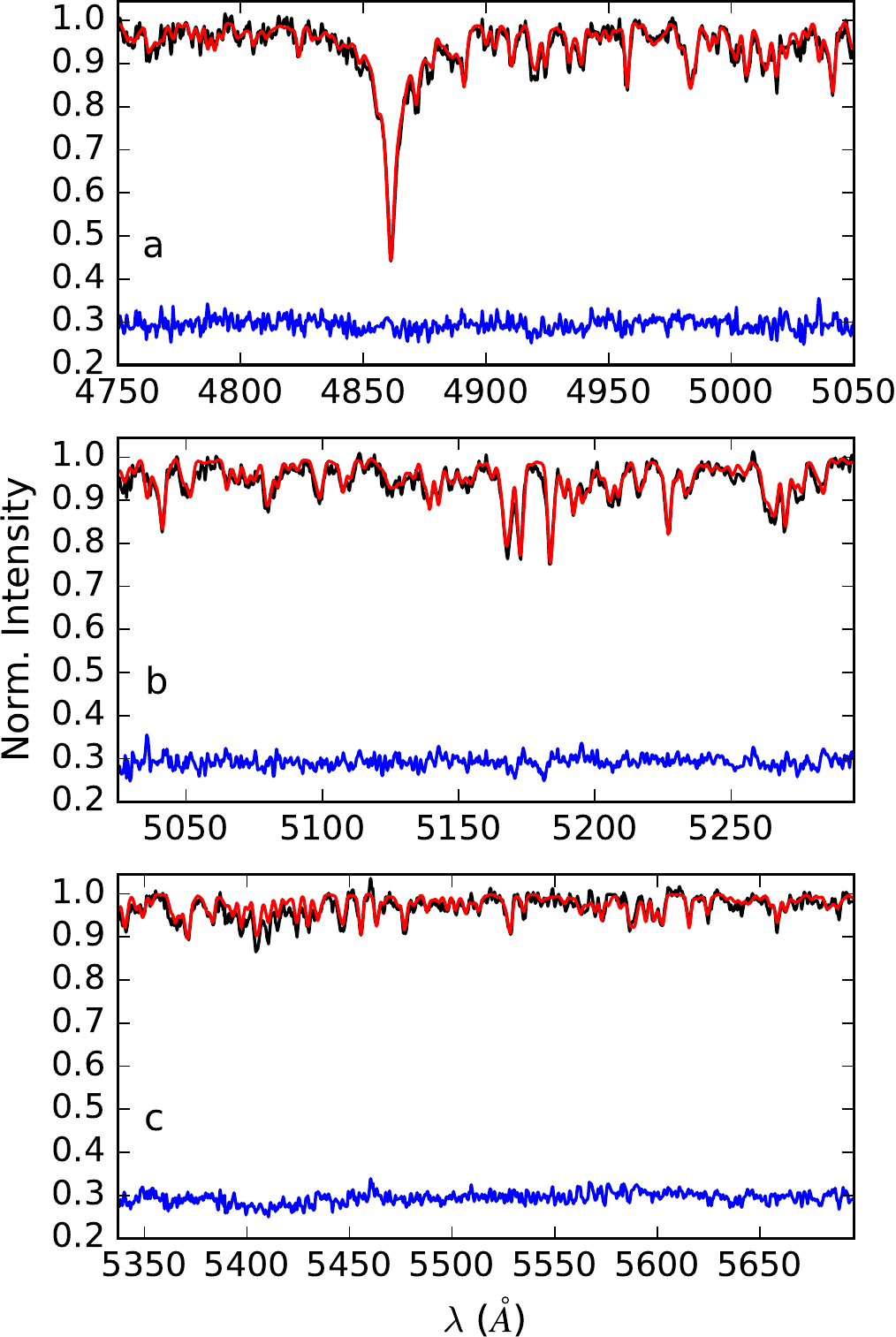}}
\caption{Representation of the observed (black), best matched (red) synthetic spectrum and 
residuals (blue) for three regions. Note that we shift the residuals upwards by 0.3 for the 
sake of simplicity. Panels $a$, $b$ and $c$ show the regions around H$_{\beta}$, Mg I triplet and 
metallic absorption lines around 5500\,\AA, respectively.}
\label{F3}
\end{figure}

Among the calculated spectra we find that the model with 6500 K effective temperature and [Fe/H]
value of $-$0.5 provides the closest match to the reference spectrum. The final effective temperature
indicates F5 spectral type \citep{Gray_2005}. Considering the effective temperature and metallicity
steps in model atmospheres, and resolution of TFOSC spectra, the final values and their estimated 
uncertainties are $T_{eff}$ = 6500$\pm$200 K and [Fe/H] = $-$0.5$\pm$0.5 dex, respectively.
Note that even we consider the neglected contribution of the secondary component in the reference 
spectrum, its effect would be fairly inside the estimated uncertainties above. The final
$T_{eff}$ values is $\sim$670 K lower than the 7166 K value estimated in \citet{Armstrong_et_al_2014}.
We show portions of reference spectrum and the model spectrum, calculated with the final parameters
above, in Figure~\ref{F3}.

%################ Figure 3 - Spectra

\subsection{Light curve modeling and physical properties}\label{S3.3}

\begin{figure}[!htb]
\centering
{\includegraphics[angle=0,scale=0.60,clip=true]{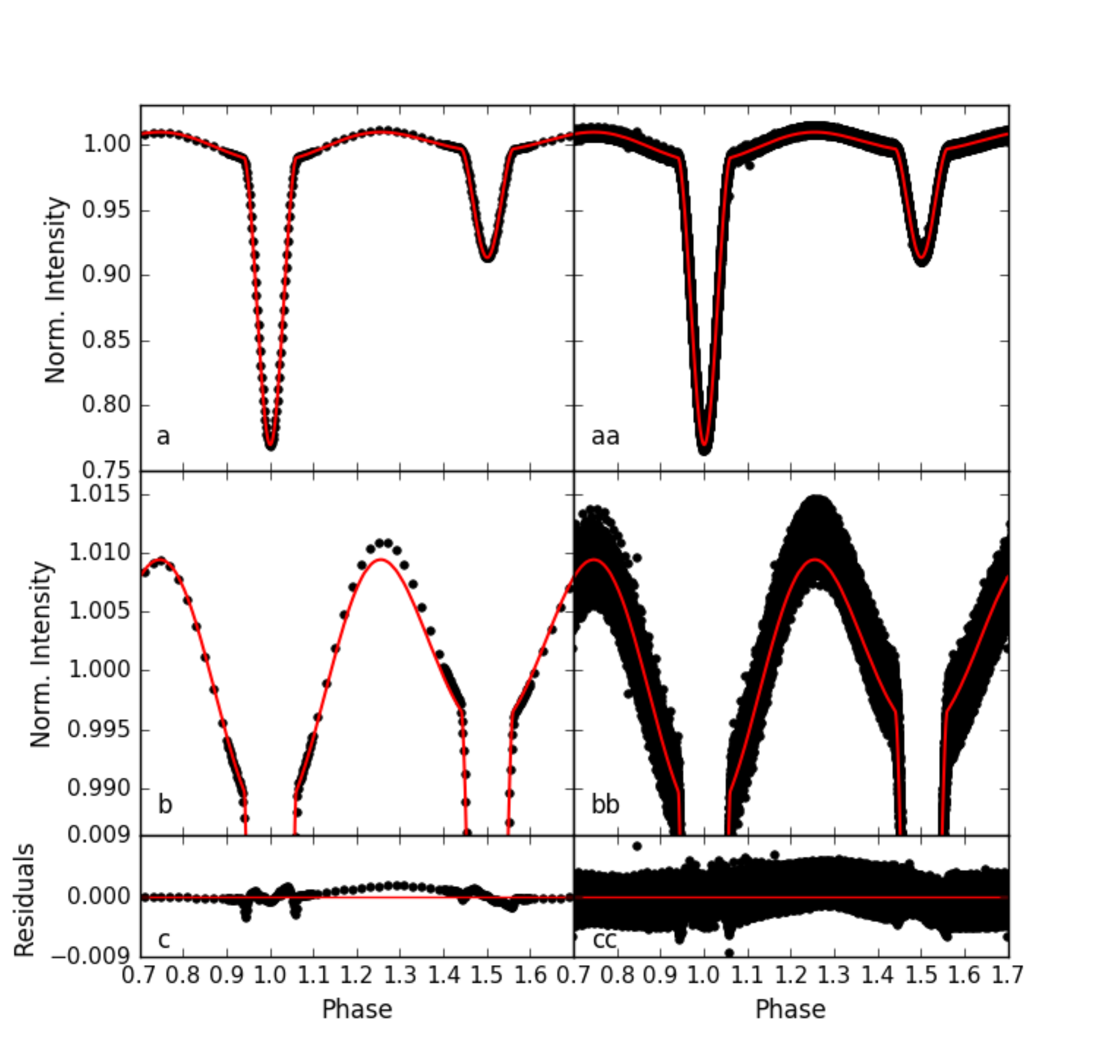}}
\caption{\textbf{a)} Phase binned light curve of KIC\,9451096 (black filled circles) 
together with best-fit model (red curves). \textbf{b)} Close up view of the light curve 
at light maxima. \textbf{c)} Residuals from the best--fit model. Panels at right 
($aa$, $bb$ and $cc$) are the same as left panels but for phased long cadence data.}
\label{F4}
\end{figure}

Global visual inspection of KIC\,9451096 $Kepler$ photometry reflects properties
of a typical close eclipsing binary. We start light curve modeling by phasing the 
whole long cadence data with respect to the ephemeris and period given by 
\citet{Borkovits_2016MNRAS_To_P}, and re--binning the phased data with a phase step of 0.002
via freely available fortran code 
$lcbin$\footnote{http://www.astro.keele.ac.uk/$\sim$jkt/codes.html$\#$lcbin} written by 
John Southworth. We plot the binned and phased light curves of the system in Figure~\ref{F4}, 
panel $a$ and $aa$. The light curve indicates detached configuration for the system. Mid-eclipse 
phases are 0.0 and 0.5 phases, indicating circular orbit. There is no conspicuous asymmetry in
the light curve.

%################ Figure 4 - LCs

We model the light curve with 2015 version of the Wilson-Devinney code \citep{WD_MAIN_1971ApJ, 
WD2015_2014ApJ}. In the modeling, we first fixed the most critical two parameters of the light 
curve modeling, i.e., mass ratio ($q$) and effective temperature of the primary component ($T_{1}$).
Since we have reliably derived these parameters in previous sections as $q$ = 0.55 and $T_{1}$ = 6500 K, 
we adopt them as fixed parameters. Calculated atmospheric properties of the primary component reveal
that both stars have convective envelopes, therefore we set albedo ($A_{1}$, $A_{2}$) and gravity
darkening ($g_{1}$, $g_{2}$) coefficients of the components to 0.5 and 0.32, respectively, which are
typical values for stars with convective outer envelopes. We also consider slight metal deficiency
of the system, thus adopt internal stellar atmosphere formulation of the Wilson-Devinney code 
according to the determined [Fe/H] value of $-$0.5. We assume that the rotation of the components 
is synchronous to the orbital motion, thus fix the rotation parameter of each component 
($F_{1}$, $F_{2}$) to 1.0. We adopt square root law \citep{Sqrt_LD_Law_Klinglesmith_1970AJ} for 
limb darkening of each component, that is more appropriate for stars cooler than 9000 K. We take 
the limb darkening coefficients for $Kepler$ passband ($x_{1}$, $x_{2}$, $y_{1}$, $y_{2}$) 
and bolometric coefficients ($x_{1bol}$, $x_{2bol}$, $y_{1bol}$, $y_{2bol}$) from 
\citet{van_Hamme_LD_1993AJ}. In the modeling, we adjust inclination of the orbit ($i$), 
temperature of the secondary component ($T_{2}$), dimensionless omega potentials of the components 
($\Omega_{1}$, $\Omega_{2}$) and luminosity of the primary component ($L_{1}$). We also include 
phase shift parameter as adjustable in the modeling since we expect a shift in ephemeris due to 
the light--time effect of the third body \citep{Borkovits_2016MNRAS_To_P}. The model 
quickly converged to a steady solution in a few iterations. We list the model output in 
Table~\ref{T3} and we plot the best--model in Figure~\ref{F4}, panel $a$, $b$, and residuals 
from the model in panel $c$.

In Figure~\ref{F4}, panel $b$, one can easily see the model inconsistency around 0.25 orbital 
phases. The inconsistency indicates an additional light variation, which is known as 
$O'Connell effect$, i.e. difference between light levels of subsequent maxima in an orbital cycle.
Possible sources of the difference may be Doppler beaming, hot spot or a cool spot on one of the
component in the system. KIC\,9451096 is a detached eclipsing binary, thus we can safely exclude
possibility of mass transfer between components, i.e. hot spot possibility. Doppler beaming was 
detected observationally among some $Kepler$ binaries \citep[see, e.g.][]{Doppler_Beaming_VanKerkwijk2010ApJ}, 
which becomes important for systems with very low mass 
ratio, especially for systems with a compact component, such as white dwarf or hot sub-dwarf. 
In addition, if the effect is in progress, then it would change light levels of each maxima. 
However, we observe inconsistency only for 0.25 phase, while the model fairly represents light 
level at 0.75 phase, thus Doppler beaming should have negligible effect in case of 
KIC\,9451096, if any. Remaining possibility is cool spots located preferably on
the cooler component.

Here we do not prefer to model this inconsistency alone, which would only show cumulative effect 
of hundreds of light curves, but instead we subtract the best--fit model from whole long cadence 
data and inspect the residuals in order to investigate further light variations. We will focus on 
this in Section~\ref{S3.4}.

% ----------------------- T3
\begin{table}[!htb]
\caption{Light curve modeling results of KIC\,9451096. $\langle r_{1}\rangle$ 
and $\langle r_{2}\rangle$ denote mean fractional radii of the primary and the 
secondary components, respectively. Internal errors of the adjusted parameters 
are given in parentheses for the last digits. Asterisk symbols in the table 
denote fixed value for the corresponding parameter. Note that we adopt the 
uncertainty of $T_{1}$ for $T_{2}$ as well, since the internal error of
$T_{2}$ is unrealistically small ($\sim$1 K).}\label{T3}
\begin{center}
\begin{tabular}{cc}
\hline\noalign{\smallskip}
Parameter & Value \\
\hline\noalign{\smallskip}
$q$ &  0.55* \\
$T_{1}(K)$ &  6500* \\
$g_{1}$, $g_{2}$  & 0.32*, 0.32*\\
$A_{1}$, $A_{2}$  & 0.5*, 0.5*\\
$F_{1}$ = $F_{2}$  & 1.0* \\
phase shift        & 0.00108(2) \\
$i~(^{\circ})$ &  79.07(4)\\
$T_{2}(K)$ &  5044(200)\\
$\Omega_{1}$ & 4.4942(49)\\
$\Omega_{2}$ & 4.8885(125) \\
$L_{1}$/($L_{1}$+$L_{2})$ & 0.897(1) \\
$x{_1bol},x{_2bol}$ & 0.136*, 0.293*\\
$y{_1bol},y{_2bol}$ & 0.583*, 0.401*\\
$x{_1}, x{_2}$  & 0.106*, 0.482* \\
$y{_1}, y{_2}$  & 0.670*, 0.313* \\
$\langle r_{1}\rangle, \langle r_{2}\rangle$ & 0.2557(3), 0.1506(5) \\
Model rms           &     3.0 $\times$ 10$^{-4}$   \\
\noalign{\smallskip}\hline
\end{tabular}
\end{center}
\end{table}

We complete light curve modeling section with calculation of absolute parameters of the system
by combining spectroscopic orbital solution and light curve model results. In Table\ref{T4}, we
give physical properties of each component. Our analysis reveals that the system is formed by
F5V primary and K2V secondary components.

% ----------------------- T4
\begin{table}
\caption{Absolute physical properties of KIC\,9451096. Error of each parameter is 
given in paranthesis for the last digits.}\label{T4}
\begin{center}
\begin{tabular}{ccc}
\hline\noalign{\smallskip}
Parameter & Primary & Secondary \\
\hline\noalign{\smallskip}
Spectral Type      &  F5V     &  K2V  \\
\multicolumn{1}{c}{[Fe/H]} & \multicolumn{2}{c}{$-0.5\pm0.5$} \\
Mass (\Msun)       &  1.18(26) & 0.65(9) \\
Radius (\Rsun)     &  1.53(10) & 0.90(6) \\
Log $L/L_{\odot}$   & 0.574(76) & $-$0.327(88) \\
log $g$ (cgs)      &  4.14(4) & 4.34(1) \\
$M_{bol}$ (mag)      & 3.31(19) & 5.57(22) \\
\noalign{\smallskip}\hline
\end{tabular}
\end{center}
\end{table}

\subsection{The out-of-eclipse variations}\label{S3.4}

\begin{figure*}[!htb]
\centering
{\includegraphics[angle=0,clip=true,scale=0.75]{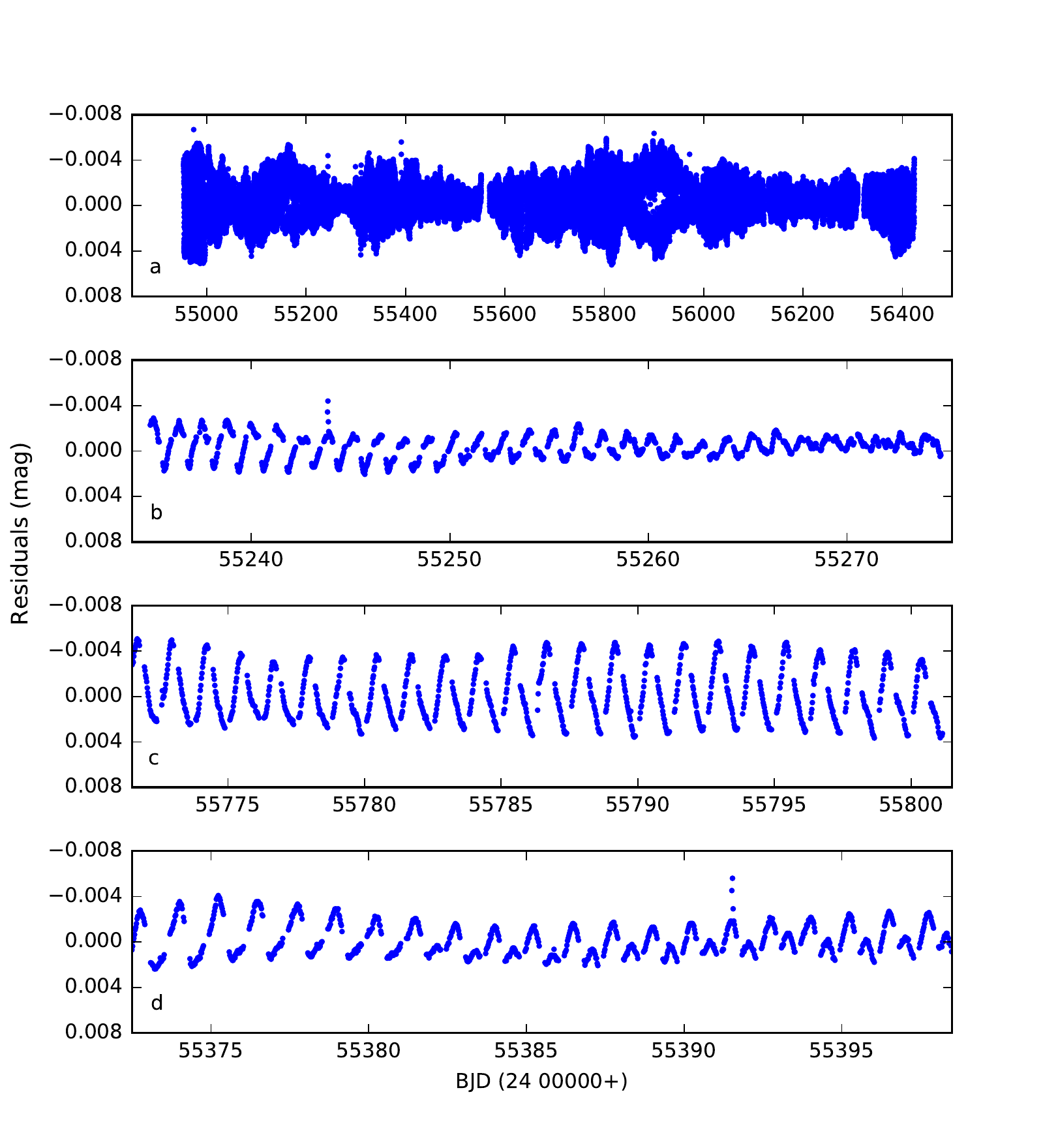}}
\caption{\textbf{a)} Residuals from whole long cadence data. Remaining panels show 
different time ranges of residuals, where we observe different light curve shapes,
and flares.}
\label{F5}
\end{figure*}

In this section, we subtract the best--fit light curve model from the whole long cadence data
and obtain residuals. Here, we first divide the whole long cadence data into subsets, where each 
subset covers only a single orbital cycle, resulting in 1026 individual light curve. Then we apply
differential corrections routine of the Wilson-Devinney code and fix all parameters, except ephemeris
reference time. In this way, we find precise ephemeris reference time for each individual subset, 
therefore eliminate any shift in the ephemeris time due to the third body reported by
\citet{Borkovits_2016MNRAS_To_P} and obtain precise residuals. In Figure\ref{F5}, we plot three 
different parts of the residuals. Note that we remove data points that correspond to the eclipse 
phases due to the insufficient representation of the model at those phases. This mainly arises 
from inadequacy of radiative physics used in light curve modeling for a very high photometric 
precision and can clearly be seen in Figure\ref{F4} panel $c$.

%################ Figure 5 - Residuals from LC

Inspecting residual brightness, we immediately see a variation pattern which changes its shape 
from time to time. Furthermore, we observe sudden increase and gradual decrease in residual 
brightness which occasionally occurs in four years of time span and has short time scale 
of a few hours. These patterns are traces of magnetic spot activity, which is very possibly from
the K2V secondary component. Observational confirmation of this possibility can be done by inspecting
magnetic activity sensitive spectral lines, such as H$_{\alpha}$ and \ion{Ca}{2} H \& K lines.
We inspect these lines in our TFOSC spectra and do not notice any emission features, which could be
considered as the sign of the activity. However, one should consider that the contribution of the
secondary component to the total light does not exceed 10\% at optical wavelengths and will steeply
decrease towards the ultraviolet region of the spectrum. Furthermore, variation patterns observed in
Figure\ref{F5} exhibit very small amplitude. Therefore the existence of magnetic spot activity can not
be confirmed or excluded via spectral line inspection in case of KIC\,9451096. Nevertheless, variation
patterns and flares observed in the residuals indicate weak magnetic spot activity on the secondary
component, which can still be detected with the very high precision of $Kepler$ photometry.

We analyze rotational modulation and flares of the secondary component via residuals by assuming
that the source of all variation patterns is only the secondary component.

\subsubsection{Photometric period and differential rotation}\label{S3.4.1}

Conventional periodogram methods for determining rotational period do not perfectly work in our 
case because observed variation patterns exhibit quick changes in amplitude and mean brightness level 
in a short time scales of a few days, which is comparable to the orbital period. Moreover, since we remove
data points at eclipse phases, this causes regular gaps in the data which repeats itself in every $\sim$0.625 
day (i.e. half of the orbital period), thus causes alias period and its harmonics, and disturbs real periods.
Furthermore, one can clearly see that the rotational modulation of residuals has asymmetric shape. Considering 
an individual light curve with an asymmetric shape, it is not possible to find a single period to represent 
whole light curve perfectly and additional periods (i.e. harmonics) are required to full representation.
Therefore we apply an alternative method based on tracing the time of a minimum light observed in an orbital cycle,
which was previously applied to RS CVn system HD\,208472 \citep{V2075Cyg_Orkun_2010AN}.
For each orbital cycle, we find the time of the deepest minimum in the cycle by fitting a second or 
third order polynomial to the data points around the expected minimum time. The order of the 
polynomial depends on the light curve shape. After obtaining all minimum times, we construct an 
$O-C$ diagram by adopting the first minimum time observed in the residuals as initial ephemeris 
reference time and orbital period as the initial period, and obtain $O-C I$ values. Then we apply 
a linear fit to the $O-C I$ values and calculate average ephemeris reference time and period given 
in Equation\ref{Eq1}, together with statistical uncertainties given in parentheses for the last digits. 

\begin{equation}\label{Eq1}
T_{0} {\rm (BJD)} = 2,454,954.02(24) + 1\fd24544(36) \ \times \ E .
\end{equation}

\begin{figure}[!htb]
\centering
{\includegraphics[angle=0,scale=0.59,clip=true]{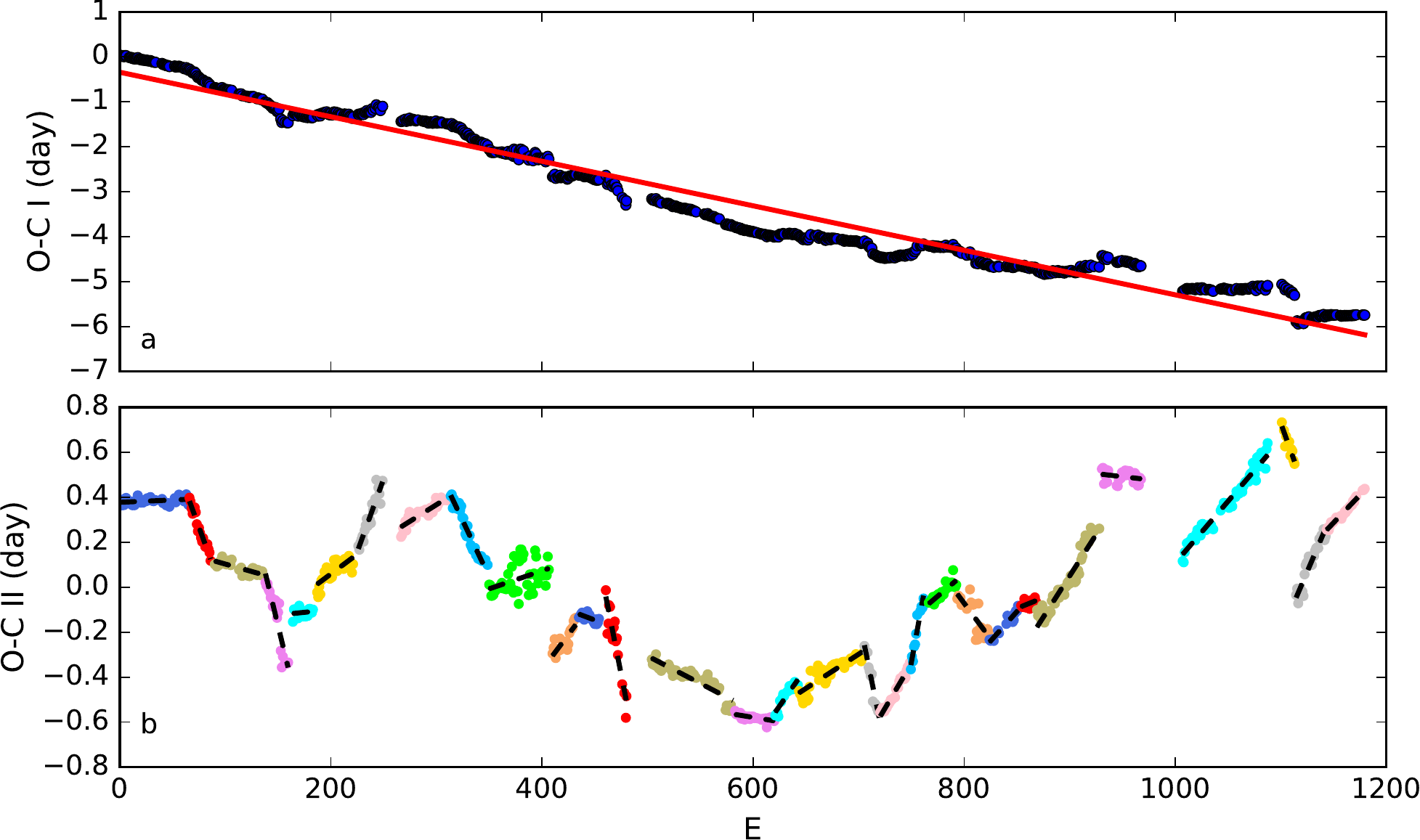}}
{\includegraphics[angle=0,scale=0.59,clip=true]{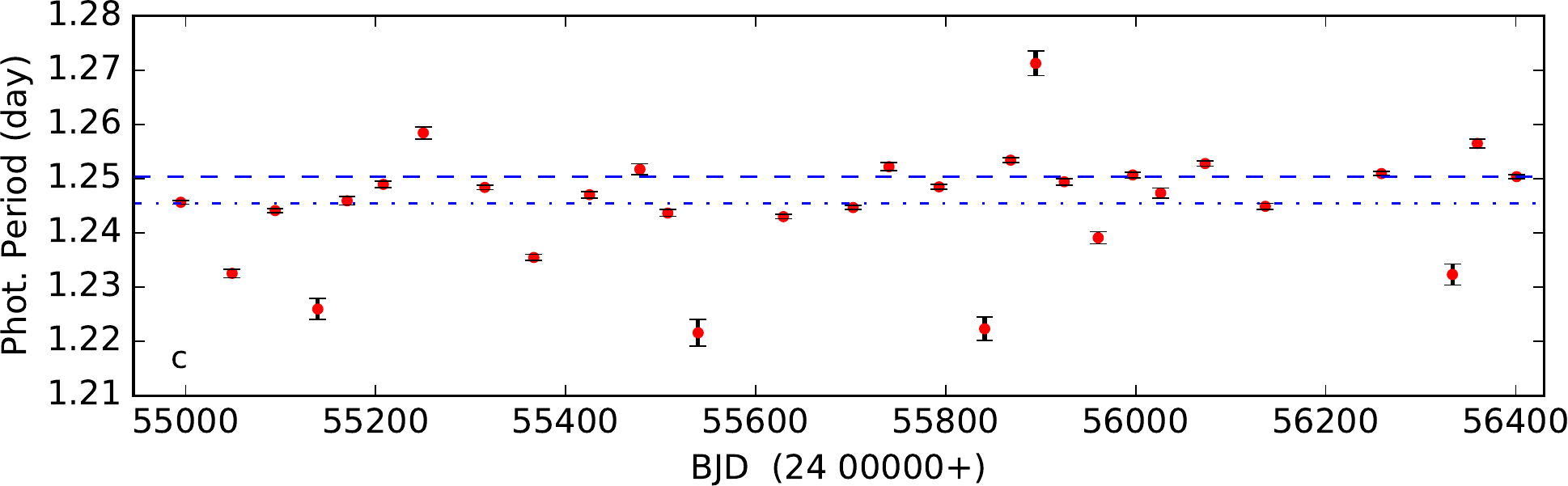}}
\caption{\textbf{a)} $O-C I$ diagram of observed minimum times (blue filled circles)
and linear fit (red line). \textbf{b)} $O-C II$ diagram obtained via residuals from the 
linear fit in panel $a$. Each color denotes a subset where data points appear on a linear
trend. Linear fit to each subset is shown by black dashed line. \textbf{c)} Calculated mean
photometric period for each subset (blue filled circles) and their statistical uncertainties. 
Note that the horizontal axis values are converted from E numbers to barycentric Julian date.
Orbital period and grand average photometric period obtained from linear fit to the $O-C I$ data
are shown with blue color in form of dashed line and dot--dashed line, respectively.}
\label{F6}
\end{figure}

In the equation, $T_{0} {\rm (BJD)}$ and $E$ denote ephemeris reference time and integer cycle 
number, respectively. We plot $O-C I$ values and linear fit in Figure~\ref{F6}, panel $a$.
After obtaining average ephemeris and period, we subtract the linear fit from $O-C I$ data 
and obtain $O-C II$ data, which in principle shows real period variation for a given time range.
Figure~\ref{F6}, panel $b$ shows $O-C II$ data. We divide $O-C II$ data into 30 subsets by 
grouping data points that appear with a linear slope. Linear trend of a subset gives the
difference between the best--fit photometric period of the subset and grand average photometric 
period given in Equation\ref{Eq1}. Therefore we can calculate mean photometric period for each
subset. We plot the calculated mean photometric periods versus time in Figure~\ref{F6}, panel $c$,
together with their statistical uncertainties. We list photometric periods for 30 subsets in 
Table~\ref{T5}, and tabulate $O-C$ analysis results in Table~\ref{T_ap}.

% ----------------------- T5 - Photometric Periods
\begin{table}
\caption{Photometric periods found from $O-C$ analysis.}\label{T5}
\begin{center}
\begin{tabular}{cccc}
\hline\noalign{\smallskip}
Subset	&	BJD	&	P	&	$\sigma$(P)	\\
	&	(24 00000+)	&	(day)	&(day)		\\
\hline\noalign{\smallskip}
1	&	54994.8107	&	1.2456	&	0.0004	\\
2	&	55048.8731	&	1.2326	&	0.0008	\\
3	&	55094.1598	&	1.2441	&	0.0004	\\
4	&	55139.0644	&	1.2260	&	0.0019	\\
5	&	55169.9192	&	1.2459	&	0.0008	\\
6	&	55208.0721	&	1.2489	&	0.0006	\\
7	&	55250.0831	&	1.2584	&	0.0011	\\
8	&	55314.8252	&	1.2484	&	0.0004	\\
9	&	55366.4562	&	1.2355	&	0.0006	\\
10	&	55425.0957	&	1.2470	&	0.0006	\\
11	&	55478.0779	&	1.2517	&	0.0010	\\
12	&	55507.4240	&	1.2437	&	0.0006	\\
13	&	55539.3828	&	1.2216	&	0.0025	\\
14	&	55629.1787	&	1.2430	&	0.0004	\\
15	&	55702.5236	&	1.2447	&	0.0004	\\
16	&	55740.2684	&	1.2522	&	0.0007	\\
17	&	55793.0150	&	1.2485	&	0.0004	\\
18	&	55840.9410	&	1.2223	&	0.0022	\\
19	&	55868.2947	&	1.2534	&	0.0005	\\
20	&	55894.6874	&	1.2712	&	0.0022	\\
21	&	55924.7567	&	1.2494	&	0.0006	\\
22	&	55960.4676	&	1.2391	&	0.0011	\\
23	&	55996.8636	&	1.2507	&	0.0005	\\
24	&	56026.2172	&	1.2474	&	0.0009	\\
25	&	56073.0738	&	1.2528	&	0.0005	\\
26	&	56136.3924	&	1.2449	&	0.0005	\\
27	&	56258.6328	&	1.2509	&	0.0004	\\
28	&	56333.3104	&	1.2323	&	0.0019	\\
29	&	56359.5423	&	1.2565	&	0.0008	\\
30	&	56400.8932	&	1.2504	&	0.0004	\\
\noalign{\smallskip}\hline
\end{tabular}
\end{center}
\end{table}

%################ Figure 6 - O-C diagram

The average period given in Equation\ref{Eq1} represents average rotation period for magnetic 
activity features on the surface of the secondary component, which are typically cool and dark 
regions, i.e., star spots, and indicates a slightly ($\sim$0.5\% day) shorter period compared 
to the orbital period. This is clearly observed in Figure~\ref{F6} panel $c$, where the 
mean photometric periods of subsets are mostly shorter than the orbital period. Assuming solar
type differential rotation, it means that the orbital period is slightly longer than the 
equatorial rotation period of the secondary component. Under the same assumption, differential
rotation coefficient can be estimated from 
$(P_{max} - P_{min})/P_{equ} = kf$, where $P_{max}$, $P_{min}$, $k$ and $f$ denote observed maximum 
and minimum period, differential rotation coefficient and a constant that depends on the range of 
spot forming latitudes, respectively \citep{Hall_Busby_1990_difrot}. Considering small 
amplitude of rotational modulation of residuals, we assume that the secondary component is not largely 
spotted and total latitudinal range of spot distribution is 45 degrees, which puts the $f$ constant 
takes values between 0.5 and 0.7 \citep{Hall_Busby_1990_difrot} Using maximum and minimum photometric 
periods from $O-C$ analysis, and assuming the shortest period corresponds to the equatorial rotation 
period of the star, we find $k = 0.081\pm0.011$ and $k = 0.058\pm0.006$ for $f = 0.5$ and $f = 0.7$, 
respectively. Since these $k$ values are calculated via boundary values of $f$, the real differential 
rotation coefficient must lie in the range of $k$ values calculated above. An average $k$ is 
found as 0.069$\pm$0.008.

\subsubsection{Flares}\label{S3.4.2}

\begin{figure}[!htb]
\centering
{\includegraphics[angle=0,scale=0.59,clip=true]{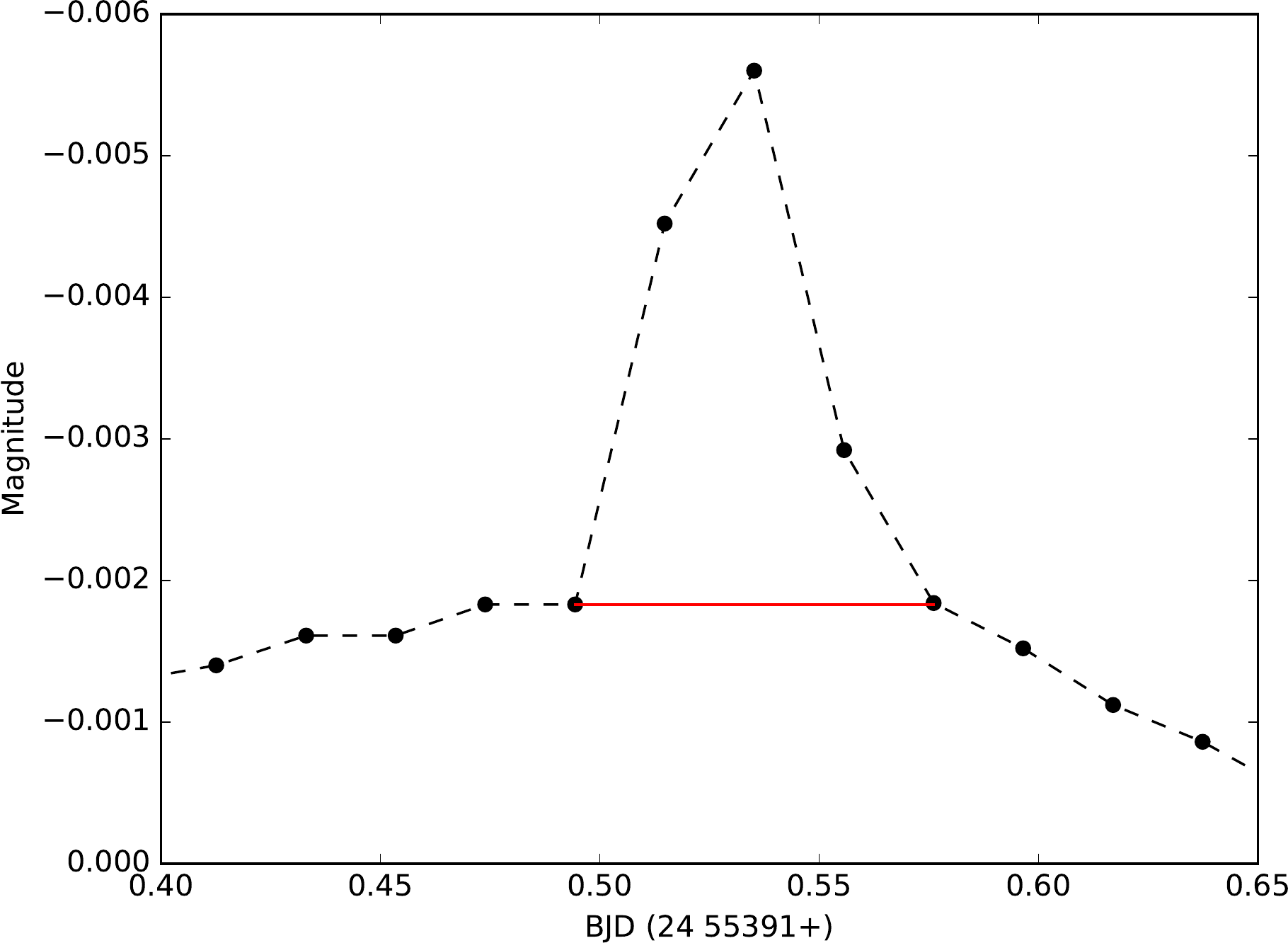}}
\caption{An example of a flare light curve. The filled black circles represent the observations, 
while the red line represents the quiescent level derived from the data out-of-flare.}
\label{F7}
\end{figure}

We detect 13 flares in the residuals from long cadence data. In flare analysis, it is critical
to determine quiescent level, which denotes the brightness level in the absence of flare.
In our case, we determine the quiescent level by applying Fourier analysis to single orbital cycle
where the flare occurs. The Fourier analysis represents the rotational modulation of residuals 
in the cycle, and then we remove the Fourier representation from the data. The remaining
residuals only shows quiescent level and flare itself. We show such a flare light curve in 
Figure~\ref{F7}.

%################ Figure 7 - Flare sample

The energy ($E$) is a very important parameter for a flare. However, the energy parameter has the 
luminosity $L$ of the star as a factor in equation $E = P \times L$ described by 
\citep{Gershberg_1972Ap&SS}. Due to the disadvantages described in \citet{Dal_Evren_2010AJ}, 
we use flare equivalent duration instead of flare energy, which is more proper. We compute the 
equivalent durations of flares via equation $P = \int[(I_{flare}-I_{0})/I_{0}]dt$ \citep{Gershberg_1972Ap&SS}, 
where P is the flare equivalent duration in seconds, $I_{0}$ is the quiescent level intensity, and
$I_{flare}$ is the intensity observed at the moment of flare. Considering the quiescent level, 
the times of flare beginning, flare maximum and flare end are determined, together with flare rise
duration, flare decay duration and flare amplitude. We list all computed values in Table\ref{T6} for 
each of 13 flare.

% ----------------------- T6 - Flare parameters
\begin{table}
\caption{The parameters calculated for each flare. Note that BJD column denotes the mid--flare
time. Tr, Td and Amp denote flare rise duration, flare decay duration and flare amplitude, 
respectively.}\label{T6}
\begin{center}
\begin{tabular}{ccccc}
\hline\noalign{\smallskip}
BJD	&	P	&	Tr	&	Td	&	Amp	\\
(24 00000+) 	&	(s)	&	(s)	&	(s)	&  (mag)\\
\hline\noalign{\smallskip}
55021.2171   &  11.4   &   1763   &   15889  &   -0.001516   \\
55043.1016   &   5.6   &   1763   &   5296   &   -0.002483   \\
55310.6569   &   7.6   &   1763   &   8830   &   -0.002047   \\
55326.5140   &   2.7   &   1771   &   1763   &   -0.001618   \\
55412.0302   &   5.9   &   1763   &   7068   &   -0.001648   \\
55416.9343   &  12.1   &   1771   &   14118  &   -0.002853   \\
55824.2162   &   4.3   &   1763   &   5296   &   -0.001578   \\
55931.1213   &   4.5   &   3534   &   3534   &   -0.001453   \\
55971.7021   &   4.9   &   1763   &   5296   &   -0.002152   \\
56142.9809   &   6.0   &   3534   &   7059   &   -0.001983   \\
56284.8887   &   3.4   &   1771   &   3525   &   -0.001806   \\
56286.5642   &   4.4   &   1771   &   3525   &   -0.001568   \\
56375.4705   &   2.2   &   1763   &   1763   &   -0.001429   \\
\noalign{\smallskip}\hline
\end{tabular}
\end{center}
\end{table}

\begin{figure}[!htb]
\centering
{\includegraphics[angle=0,scale=0.59,clip=true]{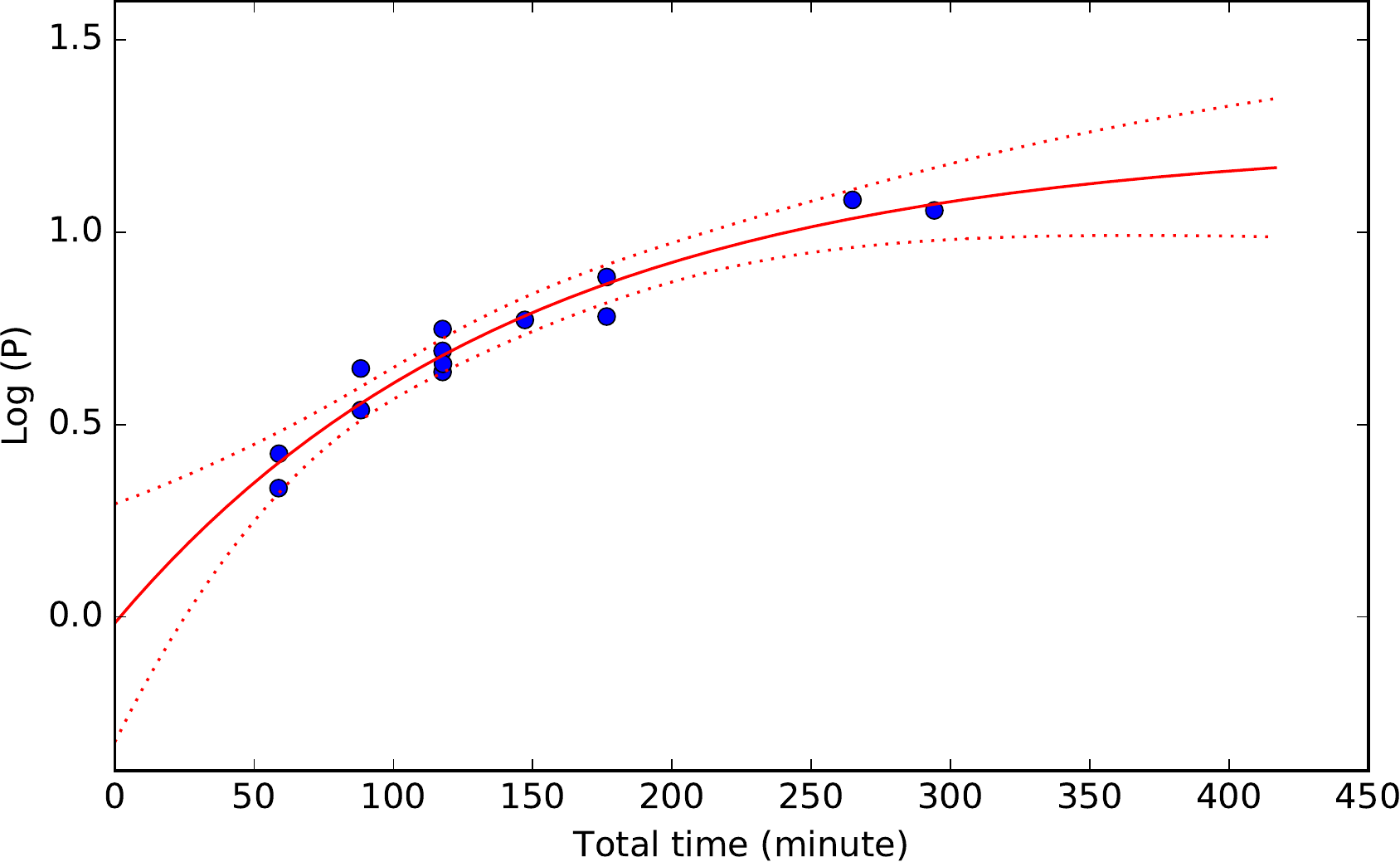}}
\caption{The OPEA model obtained over 13 flares. The blue filled circles show each flare while 
the continuous red line shows the OPEA model and the dotted red lines show the sensitivity range of 
the model.}
\label{F8}
\end{figure}

\citet{Dal_Evren_2010AJ, Dal_Evren_2011AJ} suggest that the best function to represent the
relation between flare equivalent duration and flare total durations is the OPEA, where the 
flare equivalent duration is considered on a logarithmic scale. The OPEA function is defined 
as $y = y_{0}+(Plateau-y_{0})\times(1-e^{-kx})$, where $y$ is the flare equivalent duration on 
a logarithmic scale, $x$ is the flare total duration, and $y_{0}$ is the
flare equivalent duration in the logarithmic scale for the least total duration, according to the
definition of \citet{Dal_Evren_2010AJ}. It should be noted that the $y_{0}$ does not depend 
on only flare mechanism, but also depends on the sensitivity of the optical system used in the 
mission. The most important parameter in the model is the $Plateau$ value, which defines the upper 
limit for the flare equivalent duration on a logarithmic scale and defined as saturation level 
for a star \citep{Dal_Evren_2011AJ}. Using the least squares method, the OPEA model leads to the 
results in Table~\ref{T7}. We plot the resulting model in Figure~\ref{F8} with its 95\%
statistically sensitivity limit.

% ----------------------- T7 - OPEA results
\begin{table}
\caption{Parameters derived from the OPEA Model by using the least squares method.}\label{T7}
\begin{center}
\begin{tabular}{cc}
\hline\noalign{\smallskip}
Parameter   &   Value   \\
\hline\noalign{\smallskip}
$Y_{0}$          &   $-$0.015961$\pm$0.13891   \\
Plateau     &   1.2394$\pm$0.14441   \\
K           &   0.00011438$\pm$0.000036715   \\
Half-time   &   6060   \\
$R^{2}$          &   0.94535   \\
P value     &   $\sim$0.10   \\
\noalign{\smallskip}\hline
\end{tabular}
\end{center}
\end{table}

%################ Figure 8 - OPEA model

We tested the derived model by using method proposed by \citet{D'Agostino_1986book} to understand 
whether there are any other functions to model the distribution of flare equivalent durations on 
this plane. In this method, the probability value (P value), is found to be as $\sim$0.10, which means 
that there is no other function to model the distributions \citep{Graphpad_motulsky2007, Spanier_1987}.

\citet{Ishida_1991Ap&SS} described a frequency for the stellar flare activity as 
$N_{1} = \Sigma n_{f}/\Sigma T_{t}$, where $\Sigma n_{f}$  is the total flare number detected in the
observations, while $\Sigma T_{t}$ is the total observing duration from the beginning of the 
observing season to the end. In case of KIC\,9451096 we find $N_{1}$ frequency as 0.000368411 $h^{-1}$
adopting the total long cadence observing duration as 1470.2786 days from the times of the first and 
last long cadence data points.

\section{Summary and discussion}\label{S4}

Photometric and spectroscopic analysis of KIC\,9451096 reveals that the system is composed of
a F5V primary and a K2V secondary star on a circular orbit with a detached binary configuration.
Medium resolution TFOSC spectra suggest that the system has one third of [Fe/H] of the Sun.
Light curve modeling reasonably represents observations, however, we are able to catch the 
signals of additional light variation, which is very weak compared to the variations due to the
binarity and eclipses, but still observable in the very high precision of $Kepler$ photometry.

We observe occasional flares and rotational modulation of the light curve residuals from 
eclipsing binary model. Considering the physical and atmospheric properties of the components, we 
attribute these variations to the secondary component, which is a perfect candidate for magnetic star spot
activity with its deep convective zone owing to its spectral type and very fast rotation caused 
by short orbital period. We inspect rotational modulations of the residuals to trace photometric 
period of the secondary component, and analyze its flare characteristics.

Photometric period analysis via $O-C$ diagrams shows us the average photometric period is shorter
than the orbital period by $\sim$0.5\% day. Under any type of differential rotation (either 
solar like, or anti--solar like) assumption, it means that the orbital period does not correspond
to the equatorial rotation period of the star. Following the method proposed by
\citet{Hall_Busby_1990_difrot}, we find an average differential rotation coefficient as 
$k = 0.069\pm0.008$, suggesting $\sim$3 times weaker differential rotation compared to the solar 
value of 0.19. We note that the type of differential rotation can not be 
determined from photometry alone and we implicitly assume solar type differential rotation 
in case of KIC\,9451096. However, $k = 0.069$ value, which is extracted from very high precision 
continuous photometry for a restricted time range (four years in our case), defines a lower 
limit for the strength of differential rotation on the star. Quick comparison of $k$ values
for other stars can be done by looking at 17 stars listed in \citet{Hall_Busby_1990_difrot},
where $k$ values are usually a few percent or less, except BY Dra with $k$ = 0.17.

More reliable way of detecting differential rotation with its magnitude and type is Doppler imaging, 
which is based on high resolution time series spectroscopy. Considering other stars whose $k$ values
were determined by Doppler imaging, we see mostly weak differential rotation with a $k$ value
of a few percent, either among solar type differential rotators (HD\,208472 $k$ = 0.015
\citep{DI_V2075Cyg_Ozdarcan2016}, XX\,Tri $k$ = 0.016 \citep{DI_XXTri_Kunstler_2015A&A},
$\zeta$ And $k$ = 0.055 \citep{DI_Zeta_And_Kovari2012}, KU\,Peg $k$ = 0.04 \citep{DI_KU_Peg_Kovari2016})
or anti--solar type differential rotators (UZ\,Lib $k$ = $-$0.004 \citep{DI_UZ_Lib_Vida2007AN},
$\sigma$ Gem $k$ = $-$0.04 \citep{DI_Sigma_Gem_Kovari2015}, HU\,Vir $k$ = $-$0.029 
\citep{DI_HU_Vir_Harutyunyan2016}). Due to the binary nature of KIC\,9451096, considerable
effect of tidal forces on redistribution of the angular momentum in the convective envelope
of the components can be expected, which would alter the magnitude of differential rotation
\citep{Scharlemann_tidal_difrot_1982ApJ}. Based on observational findings,
\citet{Collier_Cameron_difrot_tidal_2007AN} suggests suppression of differential rotation 
by tidal locking, which is possibly in progress for KIC\,9451096.

We detect 13 flares in residuals from long cadence data, which are attributed to the secondary
component with a corresponding $B-V$ value of 0$^m$.92 \citep{Gray_2005}. We apply OPEA model to 
analyze flare characteristic and find that the calculated flare parameters and resulting OPEA 
model parameters seem to be in agreement with parameters derived from stars analogous to the 
secondary component, except half--time value. Possible source of disagreement for half--time
value is that there are not enough sample flares at the beginning of the OPEA model.

We find $N_{1}$ value of 0.000368411 $h^{-1}$ for KIC\,9451096. $N_{1}$ was found to be 
0.41632 $h^{-1}$ for KIC\,09641031 \citep{Yol16}, 0.01351 $h^{-1}$ for KIC\,09761199 \citep{Yol17},
and 0.02726 $h^{-1}$ for Group 1 and 0.01977 $h^{-1}$ for Group 2 of KIC\,2557430 \citep{Kam17}.
Among these systems, KIC\,9451096 has the lowest $N_{1}$ value, which indicates the magnetic 
activity level of the secondary component of KIC\,9451096 is the lowest, according to 
\citet{Dal_Evren_2011AJ}.

\section*{Acknowledgments} 
We thank to T\"UB\.ITAK for a partial support in using RTT150 (Russian-Turkish 1.5-m telescope in Antalya)
with project number 14BRTT150-667. This paper includes data collected by the Kepler mission. 
Funding for the Kepler mission is provided by the NASA Science Mission Directorate. 

\bibliography{KIC9451096.bib}

\begin{appendices}

\section{$O-C$ analysis results}
We tabulate $O-C$ analysis results in Table~\ref{T_ap}. N is the number of the minimum,
beginning from the first observed minimum in the data set. $E$ is the decimal cycle number
and $E~rounded$ is the rounded E number to the nearest integer or half integer. Note that
as the time progress $O-C$ differences approach to a cycle. When this is occurred, one needs 
to add an additional increment of 0.5 to the $E~rounded$ value in order to see $O-C I$ diagram
on a trend without any discontinuity.

\begin{table*}\tiny
\caption{$O-C$ analysis results.}\label{T_ap}
\begin{center}											
\resizebox{\textwidth}{!}{%
\begin{tabular}{cccccc}											
\hline\noalign{\smallskip}											
$N$	&	BJD 	&	$E$	&	$E$	&	$O-C I$	&	$O-C II$	\\
    &  (24 00000+) &    &  $rounded$ &   (day)  &  (day)    \\
\hline\noalign{\smallskip}											
1	&	54954.3397	&	-0.01	&	0.0	&	0.037559	&	0.378738	\\
2	&	54955.5840	&	0.98	&	1.0	&	0.031503	&	0.377636	\\
3	&	54956.8162	&	1.97	&	2.0	&	0.013253	&	0.364338	\\
4	&	54958.0797	&	2.98	&	3.0	&	0.026397	&	0.382435	\\
5	&	54959.3084	&	3.96	&	4.0	&	0.004684	&	0.365675	\\
6	&	54960.5831	&	4.98	&	5.0	&	0.028962	&	0.394906	\\
\end{tabular}											
\begin{tabular}{cccccc}											
\hline\noalign{\smallskip}											
$N$	&	BJD 	&	$E$	&	$E$	&	$O-C I$	&	$O-C II$	\\
    &  (24 00000+) &    &  $rounded$ &   (day)  &  (day)    \\
\hline\noalign{\smallskip}											
76	&	55066.1529	&	89.41	&	90.0	&	-0.685052	&	0.101885	\\
77	&	55067.4011	&	90.41	&	91.0	&	-0.687308	&	0.104582	\\
78	&	55068.6338	&	91.39	&	92.0	&	-0.705006	&	0.091837	\\
79	&	55069.8833	&	92.39	&	93.0	&	-0.705847	&	0.095948	\\
80	&	55071.1320	&	93.39	&	94.0	&	-0.707572	&	0.099177	\\
81	&	55072.3928	&	94.40	&	95.0	&	-0.697178	&	0.114523	\\
\end{tabular}
}	
\end{center}
\end{table*}

\begin{table*}\tiny
\addtocounter{table}{-1}
\caption{Continued.}
\begin{center}
\resizebox{\textwidth}{!}{%
\begin{tabular}{cccccc}											
\hline\noalign{\smallskip}											
$N$	&	BJD 	&	$E$	&	$E$	&	$O-C I$	&	$O-C II$	\\
    &  (24 00000+) &    &  $rounded$ &   (day)  &  (day)    \\
\hline\noalign{\smallskip}											
151	&	55181.7781	&	181.88	&	183.0	&	-1.347004	&	-0.099451	\\
152	&	55186.8368	&	185.93	&	187.0	&	-1.289813	&	-0.022448	\\
153	&	55188.0605	&	186.90	&	188.0	&	-1.316519	&	-0.044202	\\
154	&	55189.3531	&	187.94	&	189.0	&	-1.274327	&	0.002944	\\
155	&	55190.5644	&	188.91	&	190.0	&	-1.313459	&	-0.031236	\\
156	&	55191.8713	&	189.95	&	191.0	&	-1.256925	&	0.030251	\\
\end{tabular}											
\begin{tabular}{cccccc}											
\hline\noalign{\smallskip}											
$N$	&	BJD 	&	$E$	&	$E$	&	$O-C I$	&	$O-C II$	\\
    &  (24 00000+) &    &  $rounded$ &   (day)  &  (day)    \\
\hline\noalign{\smallskip}											
226	&	55314.8625	&	288.31	&	289.5	&	-1.430004	&	0.345028	\\
227	&	55316.1085	&	289.31	&	290.5	&	-1.434445	&	0.345541	\\
228	&	55317.3359	&	290.29	&	291.5	&	-1.457361	&	0.327577	\\
229	&	55318.5720	&	291.28	&	292.5	&	-1.471718	&	0.318174	\\
230	&	55319.8259	&	292.28	&	293.5	&	-1.468206	&	0.326638	\\
231	&	55321.0985	&	293.30	&	294.5	&	-1.446001	&	0.353796	\\
\end{tabular}
}	
\end{center}
\end{table*}

\begin{table*}\tiny
\addtocounter{table}{-1}
\caption{Continued.}
\begin{center}
\resizebox{\textwidth}{!}{%
\begin{tabular}{cccccc}											
\hline\noalign{\smallskip}											
$N$	&	BJD 	&	$E$	&	$E$	&	$O-C I$	&	$O-C II$	\\
    &  (24 00000+) &    &  $rounded$ &   (day)  &  (day)    \\
\hline\noalign{\smallskip}											
301	&	55420.9879	&	373.19	&	375.0	&	-2.213674	&	-0.015172	\\
302	&	55421.7559	&	373.80	&	375.5	&	-2.070820	&	0.130159	\\
303	&	55423.4753	&	375.18	&	377.0	&	-2.227103	&	-0.018694	\\
304	&	55424.6650	&	376.13	&	378.0	&	-2.287745	&	-0.074384	\\
305	&	55425.4776	&	376.78	&	378.5	&	-2.100356	&	0.115481	\\
306	&	55426.7755	&	377.82	&	379.5	&	-2.052869	&	0.167922	\\
\end{tabular}											
\begin{tabular}{cccccc}											
\hline\noalign{\smallskip}											
$N$	&	BJD 	&	$E$	&	$E$	&	$O-C I$	&	$O-C II$	\\
    &  (24 00000+) &    &  $rounded$ &   (day)  &  (day)    \\
\hline\noalign{\smallskip}											
376	&	55532.3833	&	462.28	&	464.5	&	-2.728950	&	-0.087166	\\
377	&	55534.1421	&	463.68	&	466.0	&	-2.845732	&	-0.196520	\\
378	&	55535.3509	&	464.65	&	467.0	&	-2.887382	&	-0.233216	\\
379	&	55536.6507	&	465.69	&	468.0	&	-2.837897	&	-0.178779	\\
380	&	55537.9222	&	466.71	&	469.0	&	-2.816795	&	-0.152723	\\
381	&	55539.0799	&	467.63	&	470.0	&	-2.909571	&	-0.240547	\\
\end{tabular}
}	
\end{center}
\end{table*}

\begin{table*}\tiny
\addtocounter{table}{-1}
\caption{Continued.}
\begin{center}
\resizebox{\textwidth}{!}{%
\begin{tabular}{cccccc}											
\hline\noalign{\smallskip}											
$N$	&	BJD 	&	$E$	&	$E$	&	$O-C I$	&	$O-C II$	\\
    &  (24 00000+) &    &  $rounded$ &   (day)  &  (day)    \\
\hline\noalign{\smallskip}											
451	&	55680.7358	&	580.92	&	584.0	&	-3.799012	&	-0.565361	\\
452	&	55681.9865	&	581.92	&	585.0	&	-3.798790	&	-0.560187	\\
453	&	55683.2336	&	582.92	&	586.0	&	-3.802057	&	-0.558501	\\
454	&	55684.4673	&	583.91	&	587.0	&	-3.818739	&	-0.570230	\\
455	&	55685.7240	&	584.91	&	588.0	&	-3.812461	&	-0.558999	\\
456	&	55686.9435	&	585.89	&	589.0	&	-3.843358	&	-0.584944	\\
\end{tabular}											
\begin{tabular}{cccccc}											
\hline\noalign{\smallskip}											
$N$	&	BJD 	&	$E$	&	$E$	&	$O-C I$	&	$O-C II$	\\
    &  (24 00000+) &    &  $rounded$ &   (day)  &  (day)    \\
\hline\noalign{\smallskip}											
526	&	55784.9096	&	664.23	&	667.5	&	-4.033539	&	-0.386325	\\
527	&	55786.1128	&	665.20	&	668.5	&	-4.080768	&	-0.428600	\\
528	&	55787.3709	&	666.20	&	669.5	&	-4.073039	&	-0.415919	\\
529	&	55788.6597	&	667.23	&	670.5	&	-4.034615	&	-0.372542	\\
530	&	55789.9108	&	668.23	&	671.5	&	-4.033906	&	-0.366880	\\
531	&	55791.1424	&	669.22	&	672.5	&	-4.052701	&	-0.380722	\\
\end{tabular}
}	
\end{center}
\end{table*}

\begin{table*}\tiny
\addtocounter{table}{-1}
\caption{Continued.}
\begin{center}
\resizebox{\textwidth}{!}{%
\begin{tabular}{cccccc}											
\hline\noalign{\smallskip}											
$N$	&	BJD 	&	$E$	&	$E$	&	$O-C I$	&	$O-C II$	\\
    &  (24 00000+) &    &  $rounded$ &   (day)  &  (day)    \\
\hline\noalign{\smallskip}											
601	&	55887.0573	&	745.93	&	749.5	&	-4.418546	&	-0.365197	\\
602	&	55888.3593	&	746.97	&	750.5	&	-4.366891	&	-0.308589	\\
603	&	55889.5836	&	747.95	&	751.5	&	-4.392985	&	-0.329731	\\
604	&	55890.8943	&	749.00	&	752.5	&	-4.332688	&	-0.264480	\\
605	&	55892.1496	&	750.00	&	753.5	&	-4.327781	&	-0.254621	\\
606	&	55893.4424	&	751.03	&	754.5	&	-4.285431	&	-0.207318	\\
\end{tabular}											
\begin{tabular}{cccccc}											
\hline\noalign{\smallskip}											
$N$	&	BJD 	&	$E$	&	$E$	&	$O-C I$	&	$O-C II$	\\
    &  (24 00000+) &    &  $rounded$ &   (day)  &  (day)    \\
\hline\noalign{\smallskip}											
676	&	56011.2389	&	845.24	&	849.0	&	-4.651573	&	-0.105414	\\
677	&	56012.4835	&	846.24	&	850.0	&	-4.657366	&	-0.106255	\\
678	&	56013.7525	&	847.25	&	851.0	&	-4.638719	&	-0.082654	\\
679	&	56017.4903	&	850.24	&	854.0	&	-4.652179	&	-0.081256	\\
680	&	56018.7292	&	851.23	&	855.0	&	-4.663614	&	-0.087738	\\
681	&	56019.9859	&	852.24	&	856.0	&	-4.657367	&	-0.076538	\\
\end{tabular}
}	
\end{center}
\end{table*}

\begin{table*}\tiny
\addtocounter{table}{-1}
\caption{Continued.}
\begin{center}
\resizebox{\textwidth}{!}{%
\begin{tabular}{cccccc}											
\hline\noalign{\smallskip}											
$N$	&	BJD 	&	$E$	&	$E$	&	$O-C I$	&	$O-C II$	\\
    &  (24 00000+) &    &  $rounded$ &   (day)  &  (day)    \\
\hline\noalign{\smallskip}											
751	&	56119.5428	&	931.86	&	935.5	&	-4.507145	&	0.467436	\\
752	&	56120.8400	&	932.89	&	936.5	&	-4.460308	&	0.519226	\\
753	&	56130.7366	&	940.81	&	944.5	&	-4.566916	&	0.452241	\\
754	&	56131.9796	&	941.80	&	945.5	&	-4.574294	&	0.449816	\\
755	&	56133.2522	&	942.82	&	946.5	&	-4.552111	&	0.476952	\\
756	&	56134.5096	&	943.83	&	947.5	&	-4.545067	&	0.488948	\\
\end{tabular}											
\begin{tabular}{cccccc}											
\hline\noalign{\smallskip}											
$N$	&	BJD 	&	$E$	&	$E$	&	$O-C I$	&	$O-C II$	\\
    &  (24 00000+) &    &  $rounded$ &   (day)  &  (day)    \\
\hline\noalign{\smallskip}											
826	&	56287.7086	&	1066.35	&	1070.5	&	-5.145137	&	0.498080	\\
827	&	56288.9621	&	1067.35	&	1071.5	&	-5.142071	&	0.506100	\\
828	&	56290.2105	&	1068.35	&	1072.5	&	-5.144032	&	0.509091	\\
829	&	56291.5000	&	1069.38	&	1073.5	&	-5.104957	&	0.553119	\\
830	&	56292.7161	&	1070.35	&	1074.5	&	-5.139240	&	0.523789	\\
831	&	56293.9407	&	1071.33	&	1075.5	&	-5.164982	&	0.503000	\\
\end{tabular}
}	
\end{center}
\end{table*}

\begin{table*}\tiny
\addtocounter{table}{-1}
\caption{Continued.}
\begin{center}
\resizebox{\textwidth}{!}{%
\begin{tabular}{cccccc}											
\hline\noalign{\smallskip}											
$N$	&	BJD 	&	$E$	&	$E$	&	$O-C I$	&	$O-C II$	\\
    &  (24 00000+) &    &  $rounded$ &   (day)  &  (day)    \\
\hline\noalign{\smallskip}											
901	&	56408.3847	&	1162.86	&	1167.5	&	-5.757700	&	0.365945	\\
902	&	56409.6487	&	1163.87	&	1168.5	&	-5.744102	&	0.384496	\\
903	&	56410.8892	&	1164.86	&	1169.5	&	-5.753967	&	0.379583	\\
904	&	56412.1522	&	1165.87	&	1170.5	&	-5.741409	&	0.397095	\\
905	&	56413.4054	&	1166.87	&	1171.5	&	-5.738612	&	0.404844	\\
906	&	56420.9048	&	1172.87	&	1177.5	&	-5.741557	&	0.431616	\\
907	&	56422.1512	&	1173.87	&	1178.5	&	-5.745583	&	0.432544	\\
908	&	56423.4023	&	1174.87	&	1179.5	&	-5.744909	&	0.438170	\\
\noalign{\smallskip}\hline											
\end{tabular}
}										
\end{center}											
\end{table*}

\end{appendices}

%\begin{thebibliography}
%\bibitem[Arthur \& Hoare(2006)]{2005astro.ph.11035A} Arthur, S.~J., \&
%Hoare, M.~G.\ 2006, \apj, in press (astro-ph/0511035)
%\end{thebibliography}

\end{document}